# Self-Organization and the Physics of Glassy Networks

P. Boolchand[1], G. Lucovsky[2], J. C. Phillips[3], and M. F. Thorpe[4]

1. Department of ECECS, University of Cincinnati, Cincinnati, OH, 45221-0030
2. Department of Physics, North Carolina State University, Raleigh, NC, 27695-8202
3. Dept. of Physics and Astronomy, Rutgers University, Piscataway, N J, 08854-8019
4. Dept. of Physics and Astronomy, Arizona State University, Tempe, AZ., 85287-1504

ABSTRACT

Network glasses are the physical prototype for many self-organized systems, ranging from proteins to computer science. Conventional theories of gases, liquids, and crystals do not account for the strongly material-selective character of the glass-forming tendency, the phase diagrams of glasses, or their optimizable properties. A new topological theory, only 25 years old, has succeeded where conventional theories have failed. It shows that (probably all slowly quenched) glasses, including network glasses, are the result of the combined effects of a few simple mechanisms. These glass-forming mechanisms are topological in nature, and have already been identified for several important glasses, including chalcogenide alloys, silicates (window glass, computer chips), and proteins.

"Some four thousand years ago man discovered glass in the embers of a fire built somewhere in the deserts of the Near East. A few years from now ultratransparent glass fibers will be transmitting [much, much, much!] more information than copper wires" [1]. As foreseen 25 years ago, the Internet has indeed revolutionized our world. In this article certain phrases will appear in {}s: the interactive reader will find background on the quoted topics with his Internet Browser. What was unforeseeable 25 years ago was the discovery of an important new class of *electronic* glasses, doped cuprates, exhibiting not only high temperature superconductivity (~ 100K), but also "strange metal" transport anomalies right up to the annealing temperature (700 K) – a sure sign of electronic glassy arrest. Theory has identified two new topological phase transitions in network glasses, and a new topological phase characterized by self organization; it appears that this new phase is also responsible for high temperature superconductivity in the cuprates. Finally, there is an exciting new bridge between physics and biophysics, based on analyzing strain in protein networks using methods parallel to those used for strain in network glasses.

Glass is so much a part of our daily lives that we take its properties for granted. The most important glasses (like window glass) have aperiodic, yet microscopically homogeneous, network structures. They are formed by quenching to temperatures far below the equilibrium melting point, and they can remain in their metastable states indefinitely. Physical glasses – not just network glasses, but also molecular glasses, commercial polymers, the optimized electronic networks of cuprates, and collapsed folded peptide chains of proteins – all share an essential property: they are {space-filling} (or nearly so) networks. Mathematicians have studied {space filling} by {sphere or polyhedral packing} in great detail, but the way in which network glasses are connected to fill space is different (much more abstract), and the differences are very important in practice.

The origin of these differences lies typically in skeletal constraints that define the network configuration. Fig. 1 shows examples of such constraints, here represented by nearest neighbor springs for an N = 4-membered cluster in d = 2 dimensions. In real glasses constraints are imposed not just by nearest neighbor contacts (bond stretching) – here called σ (strong) forces, as occurs for sphere packing – but also for π forces (bond-bending, weaker three-body forces). The radius of a cluster of N atoms increases like $N^{1/3}$, and the forces decrease with distance, but at the same time, the number of possible cluster configurations increases ~ $e^N$, leading to statistical problems of exponential complexity. These topological problems are well known to mathematicians, who describe them as "Not Polynomial (NP) complete" (We prefer {exponentially complex}.) A familiar example of an {exponentially complex} problem is the traveling salesman, who should visit randomly placed cities along the shortest path; many combinatorial problems are {exponentially complex}. There are now computer algorithms that have solved problems of 25,000 cities in "only" 70 computer years (parallel processing). However, these graphic methods do not include the physical restriction that the cities form a space-filling megapolitan network, and so they cannot be extended to network glasses or proteins.

The exponentially great simplification brought about by the space-filling network restriction is embodied [1] in a very simple yet extremely far-reaching idea: The number of interatomic Lagrangian constraints (he and we count the number of potentials, *not* the number of Newtonian forces) $N_c$ that are intact at the glass transition temperature $T_g$ should equal the number of degrees of freedom $N_d$. There are three possible cases (Fig. 1). If $N_c < N_d$, the underconstrained glass is "floppy", and it could freely crystallize, while if $N_c > N_d$, the glass would be rigid and stressed, and could relieve the stress by crystallizing exothermically (in practice, severely overconstrained materials like amorphous Si films crystallize explosively when heated). When $N_c = N_d$, the network is rigid yet stress-free. In this ideal case (called "isostatic"), the dynamics of the glass are arrested because it is trapped "in limbo" [1]. Intact constraints can be identified in diffraction or optical data as anomalously narrow peaks. But is this idea really correct? Does it "miraculously" identify the optimally space-filling, exponentially small fraction of configuration space that is glassy?

It does quite well (enough precision (~ 1%) for practical purposes, as confirmed by many experiments), and Fig. 2(a) shows the quantitative results [2]. One can construct space-filling tetrahedrally coordinated models of amorphous silicon, disordered in various ways, and then dilute the connectivity of the network by striking out bonds at random until the average coordination number has been reduced from four to nearly two. At this level the network has begun to fall apart into disconnected chain segments, and there are many cyclical (zero frequency) vibrational modes. In reality weak additional forces will give these modes a small finite frequency of a few wave-numbers and so we will refer to them as floppy modes. One now adds cross-linking bonds at random, and the number of such floppy modes decreases. Remarkably, that decrease is nearly linear in average coordination number $<r>$, and the latter itself is linear in the number of intact constraints; it extrapolates to zero at the mean-field value of 2.40 = 12/5 (calculated theoretically from σ and π constraints). More sophisticated numerical simulations have shifted the transition to 2.39.

Cyclical, or Goldstone, modes play an important role in theoretical physics: they are usually explained as the result of some kind of symmetry. However, this disordered network has no obvious symmetries, and if one examines the vibrational matrix, and counts the number of off-diagonal Newtonian forces generated by the three-body potentials, it is much too large to explain the large number of cyclical or floppy modes for <r> < 2.40. This "counter-intuitive" behavior is dramatic evidence for the power of Lagrange's concept of constraints (he and we count the number of potentials, not forces), here applied in the statistical limit where both $N_c$ and $N_d$ are large, but the fractional difference between them is small. It is fair to say that this idea, as confirmed in Fig. 1, has revolutionized our understanding of the microscopic nature of glass and other networks of atoms. This idea has sweeping implications: for instance, it shows that all the very common central force (sphere packing) models of glasses are non-trivially incomplete.

The first dramatic experimental confirmation of the power of the "limbo" concept came from Raman and Mossbauer experiments [3] on chalcogenide glass alloys (S and Se cross-linked by As, Ge, Si, Sn,…). These experiments examined the bond and site connectivity of the alloys and revealed that some kind of phase transition was occurring near <r> = 2.40. An example is shown in Fig. 3 for (Sn, Ge)Se alloys. Replacing Ge (high melting point) with Sn (low melting point) dilutes the π bending constraints, and the limbo composition is calculated to occur at 40% Sn, which is exactly where $T_g$ collapses. Mossbauer site populations exhibit parallel behavior.

Subsequent work repeatedly confirmed the existence of the stiffness transition at or very near <r> = 2.40; later, however, it suggested that there could actually be *two* transitions, rather than just the one predicted by mean-field theory for a "random" space-filling network. Between the two transitions is a new glassy phase, the "intermediate phase", which is strictly a topological phase. The phase has no special symmetry, but it does have a characteristic connectivity. It is a percolative phase, but the percolation is not occurring on a lattice, and it is not random. At first one might think that non-equilibrium "off-lattice" percolation behaves identically to equilibrium "on-lattice" percolation, but this is apparently not the case. In a sense this is obvious, as there are two non-equilibrium stiffness transitions, which bound the non-equilibrium intermediate phase, and only one equilibrium critical point, with no intermediate phase. Of course, because this glassy phase is not an equilibrium phase, it is not a true phase in the sense of equilibrium thermodynamics. However, its boundaries with respect to composition are often sharp, justifying the terminology.

Some experiments have identified an electronic intermediate phase, and the electronic cases often generate disputes involving complex quantum interactions. However, it appears that the same intermediate phase occurs in insulating network glass alloys, where quantum effects are secondary; in the latter the topological nature of the intermediate phase is unambiguous. The generic similarities between these intermediate phases are striking, and it seems quite probable that they all share a common topological origin.

Specific heat experiments on glasses have traditionally involved a technique called differential scanning calorimetry (DSC), but when these are modified to include modulation with lock-in detection, one gains in two ways: accuracy is increased by an order of magnitude, and one now measures both the reversible and the non-reversible enthalpy of the glass transition. The new instruments are widely used in industry for quality control, especially of plastics, but are entering research practice only slowly. Fig. 4 shows dramatic results for several glass transitions [4]. The nonreversible enthalpy of the transition varies extremely rapidly near the two stiffness transitions, dropping in favorable cases nearly to zero in the intermediate phase, and forming a "reversibility window". (Of course, all glass transitions are irreversible; this term is merely a convenient way to describe the very deep recently discovered valley, whose existence had previously not even been suspected.) Also shown is the much shallower (and unremarked) valley in a more traditional quantity, the logarithmic derivative of the viscosity in the melt, extrapolated to $T_g$, said to measure the "strength" of the glass (loosely speaking, its resistance to crystallization, which decreases only by a factor of 2 in the window). The data base for chalcogenide alloy glasses includes many different compositions (20-30 for an alloy series) for many alloys (about ten). Overall the window spans a range of <r> from 2.28 to 2.52, and is (probably somewhat fortuitously) centered on 2.40(3). Clearly the reversibility window defines a glassy "phase" that will have physical properties quite different from those of glasses with compositions outside the window.

The reversibility window is only one of the remarkable properties of the intermediate phase. It is also often a maximal *plateau* (not peak!) in molar density. (In other words, the result of the space filling condition is at first an increase in density when chains are cross-linked, followed by a decrease when the network is overconstrained. However, in the strain-free intermediate phase, the density is nearly constant as percolative strain-free backbones are added to the network. For this reason the intermediate phase can be said to be in isostatic mechanical equilibrium; isostatic is a term borrowed from hydrodynamics to describe the absence of nonlinear strain effects in the intermediate phase.) When one monitors the aging or relaxation of the glassy network by changes in the non-reversible enthalpy of the glass transition over periods of order years (see Fig. 5), there is almost no aging in the intermediate phase [5] and substantial aging outside it, a fact of great importance for quality control.

The discrete methods of set theory and topology used to describe glassy networks are also used in computer science. The parallels are more than casual: a "phase transition" occurs between computable and non-computable problems which closely resembles the stiffness transition, a discovery that followed quickly on the heels of the discovery of the intermediate phase. Moreover, again there is an intermediate phase of intermediate (that is, possible, but very difficult) computability [6].

The discovery of a new inorganic phase in glasses, with a nearly reversible, nearly non-aging character, opens new vistas for understanding non-equilibrium behavior. Internal strain fields govern this behavior; traditionally such off-lattice strain fields have defied our best efforts to achieve quantitative understanding. They are not only long-range, but also non-local. They are easy to understand in crystals, where the lattice framework makes it easy to treat elastic misfit and {misfit dislocations}. Are there other applications of these very general concepts?

Consider window glass first; it may well be the material most responsible for modern civilization. Window glass is based on silica, but the melting point of $SiO_2$ is much too high for practical processing. So soda ($Na_2O$) is added to reduce the melting point at the eutectic, but the soda creates non-bridging oxygen dangling alkali ends (-Si-O-Na) that are reactive. Empirically it was found that adding lime (CaO) zips up these ends and makes window glass chemically stable. Can topology calculate how much soda (x) and lime (y) should be added to silica (1-x-y) to form window glass? It can! Two conditions are needed to determine x and y. The first condition is global: the strain-free limbo condition, while the second condition is local (the average number of rings passing through a cation should be the same as in silica, namely, six). These two conditions are illustrated by the ternary phase diagram (Fig. 6); they give x = 0.16 and y = 0.10, in perfect agreement with commercial practice [7].

While window glass is very old, the next example is essential to modern electronics. Trick question: why is it that computer chips are usually made from Si, rather than Ge? Ge has many electronic advantages, including higher carrier mobilities, lower Joule heating, etc. The reason is that electronic factors are usually secondary to a crucial materials factor, the perfection of the native oxide interface, $M:MO_2$. For M = Si this interface is nearly ideal, and the question is, how and why does crystalline Si form a nearly ideal interface with glassy $SiO_2$? The ideality of the Si-$SiO_2$ interface corresponds to a defect level near $10^{-4}$, while simple misfit considerations (similar to those used for crystalline interfaces) would lead one to expect defects levels of 10%.

{High k dielectrics} (silica films doped with Zr or Hf) are now technologically interesting: what are their minimal defect densities?. Topology explains the perfection of the undoped interface in terms of an SiO monolayer; it turns out that when the constraints are calculated carefully, and allowance is made for breaking of π constraints involving non-resonant O-Si-Si interlayer bonds, that this monolayer is actually isostatic, that is, strain-free, which means it can be ideal. Moreover, the theory [8] enables one to identify the relevant chemical variables for both the defect (charge trapping) density (Fig. 7), which is the main factor limiting transistor speed, and the dielectric constants of the doped films (Fig. 8).

The electronic applications of topological network theory are very interesting, but they involve a variety of quantum factors that the interested reader can study elsewhere [9]. There are a few points of general significance. Efforts are often made to describe the metal-insulator transition in semiconductor impurity bands, the intermediate phase of high temperature superconductors and sometimes even the molecular glass transition, in terms of a *single* continuum phase transition and its associated critical point, but experiment does not support these efforts based on equilibrium models. There are definitely two well-separated impurity band transitions, one electronic and one thermal, and there is an intermediate phase between them with scaling properties different from those predicted by continuum critical point theory. In the cuprates all the experimental evidence, including the isotope effect, supports a picture of three-dimensional percolation along topological paths determined by minimizing strain and maximizing dielectric screening energies associated with dopants located *outside* the ubiquitous $CuO_2$ planes. The latter are crucial primarily for architectonic, not electronic reasons; they are isostatic planar networks (much like the SiO monolayers discussed above), and they stabilize a soft network that would be otherwise insulating because of the {Jahn-Teller} effect.

Because better samples are now available, as well as synchrotron beams with higher intensity and better resolution, there is now a growing wave of evidence [10] for strong electron-phonon interactions in the cuprates; these also reflect enhanced interactions of carriers at the Fermi energy with disordered (glassy) dopants. This suggests that topological models of soft hosts and glassy dopants will play an increasingly important part in developing genuine understanding of the physics of high temperature superconductivity. Many studies of older intermetallic superconductors (such as NbN and $Nb_3Sn$) showed that the same strong electron-phonon interactions that produce high temperature superconductivity also cause lattice instabilities that limit the attainable temperatures. Just as these limits are reached, one finds anomalies not only in vibrational dispersion curves, but even in the lattice constants themselves. The most striking recent data [11] concern the effect on dispersion of quasiparticle states near the Fermi energy on replacing $^{16}O$ with $^{18}O$ (see Fig. 9). Because of glassy arrest of the dopant configuration during sample annealing, the observed isotope effects on phonon fine structure in the cuprates show many anomalies.

Self organized backbones immersed in soft matrices, even in space-filling inorganic networks, are strongly suggestive of the native states of proteins in water, collapsed by a combination of hydrophobic interactions in their cores and hydrophilic interactions nearer their surfaces [12]. Fig. 2(b) shows the strong similarities in network topologies of proteins and glasses. Protein structures exhibit a high density of local close packing, characteristic of densely packed spheres, involving the side groups of the peptide chains.

Sphere-packing models [13], of course, include only σ (main chain) interactions, whereas the dynamical properties of proteins near their transition states must be determined by primarily π (bending or shear) interactions. This suggests using hydrogen bonds and side group (amino acid) contacts to index protein structures near their transition states [14]. One begins with a mechanical model of the completely collapsed, largely dry static protein structure determined by diffraction and molecular dynamics (the “native” state), and dilutes the connectivity by cutting hydrogen bonds and hydrophobic “tethers” starting with the weakest. In this way one arrives at the transition state, defined as a maximum in the free energy and the state at which non-trivial floppy modes first occur in significant numbers. (Just as with glasses (Fig. 2), this state occurs very close to an average coordination number of 2.40, which provides a model-independent topological *definition* of the transition state.)

This leads us to a paradox; paradoxes often occur in exponentially complex problems. Native structures are determined by combining structural information from diffraction with relaxation of the structure according to molecular dynamics potentials. In particular, the diffraction data are resolution limited, and generally do not determine hydrogen bonding configurations. Thus a dynamical simulation technique is used to obtain the full *static* structure. However, for typical proteins containing 100 or more amino acids, the simulations of protein *dynamics* cannot be carried out for much more than tens of nanoseconds, whereas in reality protein (or glass transition) dynamics usually evolve over milliseconds. More generally, such difficulties are *characteristic* of {exponentially complex} physics, and they can be used to "prove" that proteins cannot have evolved without some kind of special help: for a typical protein with N ~ 100 amino acids, there are ~ $10^{300}$ possible combinations of the sequences formed from 20 amino acids that determine the protein; this number of possible combinations could not have evolved haphazardly. (Biochemists call this {Levinthal's paradox}.)

What is this special help that made the creation of proteins possible? In the intermediate state, which is presumably the state in which proteins are formed, the peptide chains are strain-free. Mathematicians in a subfield called "graph theory of rigidity" have studied the flexibility and rigidity of mechanical networks, following the first work of Maxwell (1863) on scaffolds. Their work shows that when two strain-free networks are interlaced, the combined network is also strain free. This simple additivity or closure condition has a startling effect on the combinatorial complexity of proteins: instead of being exponential in the number N of peptide units in the protein, the complexity is probably ~ $N^2$, but nearly linear in N practically. Thus placing protein dynamics, including protein formation, in the intermediate (strain-free) phase resolves {Levinthal's paradox} in principle, even though there may still be questions about the reliability of native structures and hydrogen bonds for any particular case. It also explains the origin of the mysterious and not easily quantified{folding funnels}that are popularly used to explain the evolutionary origin of protein functionality and reversibility [15].

A special algorithm [16], based on graph theory, called the {pebble game} (similar games are popular in computer science for solving hierarchical problems) identifies the rigid or overconstrained regions of a mechanical network and separates them from the soft regions. This leads to realistic pictures of the dynamical behavior of proteins (such as the protein HIV protease shown in Fig. 10). Because of the Lagrangian simplifications inherent in restricting the structure to the intermediate phase, the calculations required to obtain such pictures take only minutes on a PC, even for very large proteins. Main-chain flexibility appears also to be critical to protein – ligand (drug) interactions (called docking), as illustrated in a study of binding of a candidate 17 Carbon "drug" molecule to a 165 amino acid protein [17].

There have been many workshops and conferences devoted to the "unconventional" topics discussed here [18,19].

## REFERENCES

1. J. C. Phillips, Physics Today **35** (2) 27 (1982).
2. H. He and M. F. Thorpe, Phys. Rev. Lett. **54** 2107 (1985).
3. M. Stevens, J. Grothaus, P. Boolchand and J.G. Hernandez, Solid State Comm. **47**, 199 (1983).
4. Y. Wang, J. Wells, D. G. Georgiev, P. Boolchand, K. Jackson, and M. Micoulaut, Phys. Rev. Lett. **87** 185503 (2001).
5. U. Vempati and P. Boolchand, J. Phys. Cond. Mat. **16** S5121 (2004).
6. M. Mezard and R. Zecchina, Phys. Rev. E **66** 056126 (2002).
7. R. Kerner and J. C. Phillips, Solid State Comm. **117** 47 (2001).
8. G. Lucovsky and J. C. Phillips, App. Phys. A – Mat. Sci. **78** 453 (2004).
9. . J. C. Phillips, Phys. Rev. Lett. **90** 059704 (2003).
10. V. L. Ginzburg and E. G. Maksimov, Phys. Stat. Sol. 242, 9 (2005).
11. G.-H. Gweon *et al.*, *Nature*, **430** 187 (2004).
12. A. R. Fersht, *Structure and Mechanism in Protein Science: A Guide to Enzyme Catalysis and Protein Folding* (Freeman, New York, 1999).
13. A. Soyer, J. Chomilier, J.-P Mornon, R. Jullien, and J. F. Sadoc, Phys. Rev. Lett. **85** 3532 (2000).

14. A. J. Rader, B. M. Hespenheide, L. A. Kuhn, and M. F. Thorpe, Proc. Nat. Acad. Sci. (USA) **99** 3540 (2002).

15. P. G. Wolynes, *Proc. Nat. Acad. Sci.* **101** 6837 (2004)

16. D. J. Jacobs and. M. F. Thorpe, Phys. Rev. Lett. **75** 4051 (1996).

17. M. I. Zavodsky, M. Lei, M. F. Thorpe, A.R. Day, and L. A. Kuhn, Proteins: Structure, Function and Bioinformatics **57** 243–261 (2004) .

18. *Rigidity Theory and Applications* (Ed. M. F. Thorpe and P. Duxbury, Kluwer Academic/Plenum Publishers, New York 1999).

19. *Phase Transitions and Self-Organization in Electronic and Molecular Networks* ((Ed. M. F. Thorpe and J. C. Phillips, Kluwer Academic/Plenum Publishers, New York, 2001).

## Figures and Captions

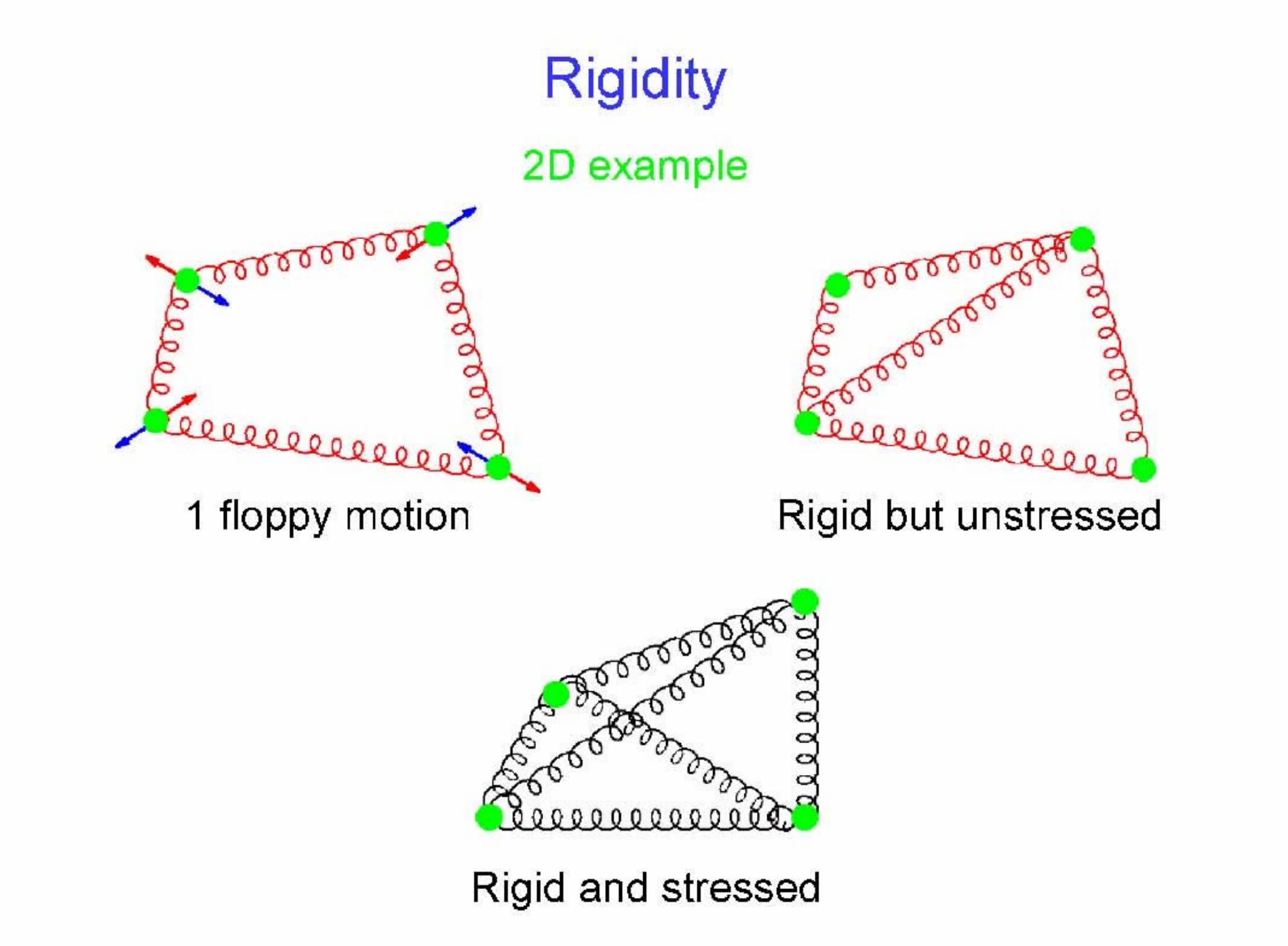

Fig. 1. Three examples of elastic network skeletons in two dimensions with nearest neighbor forces only. A quadrilateral is floppy (isostatic, stressed) with no (one, two) cross braces. Note that there are eight degrees of freedom for the four atoms, but three of these are associated with two translational and one rotational rigid modes. Each spring here is an independent constraint.

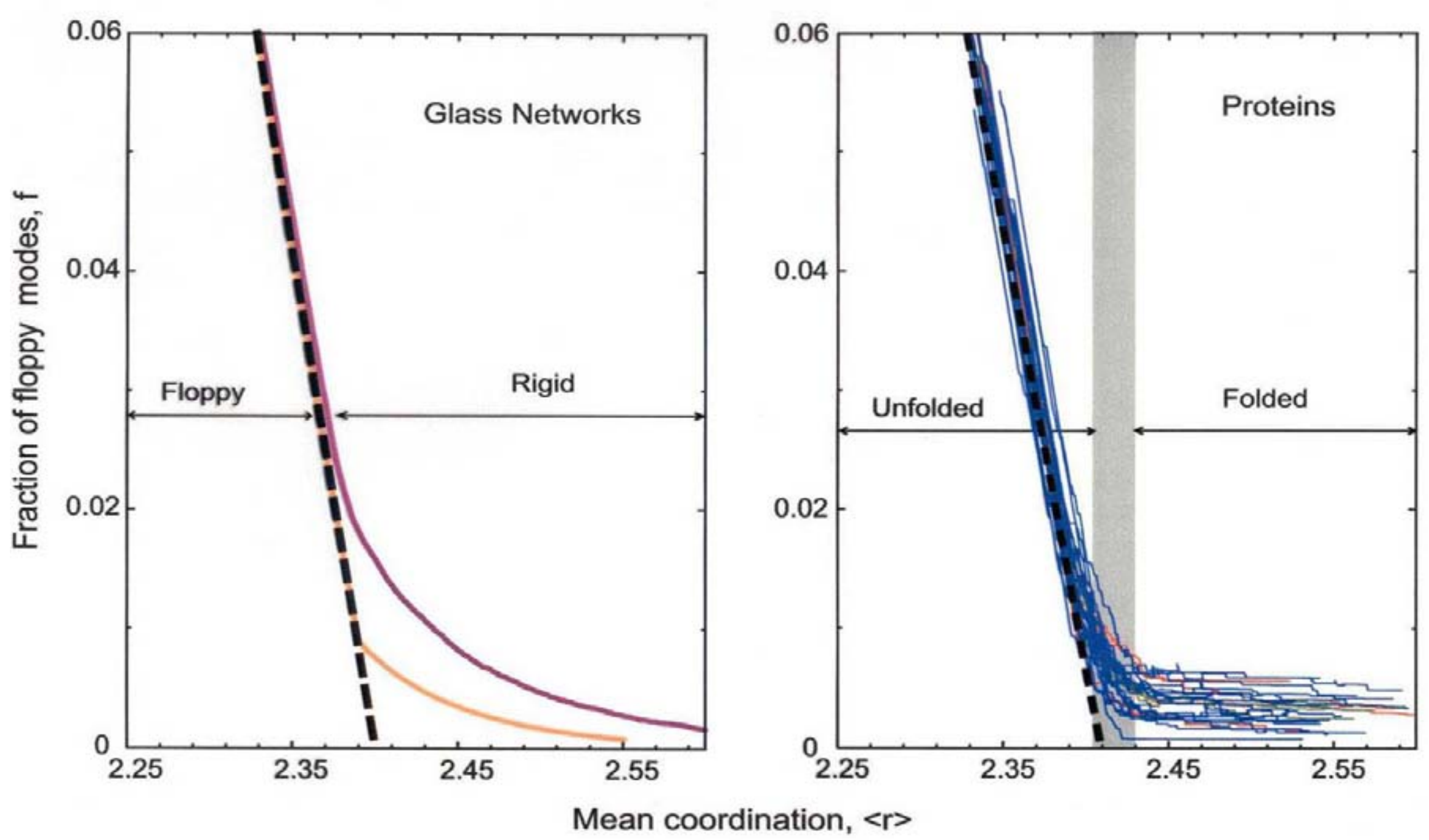

Fig. 2 (a) The number of floppy (cyclical) vibrational modes per degree of freedom for a diluted diamond lattice with several thousand atoms. The random case gives a smooth curve, while allowance for the formation of percolative self-organized backbones suggests a first-order transition from floppy to rigid.(b) A similar plot for 20 proteins, each also involving several thousand atoms. The similarity is used to help calibrate the relative strengths of hydrogen bonds and hydrophobic interactions, and to identify the transition states of the proteins.

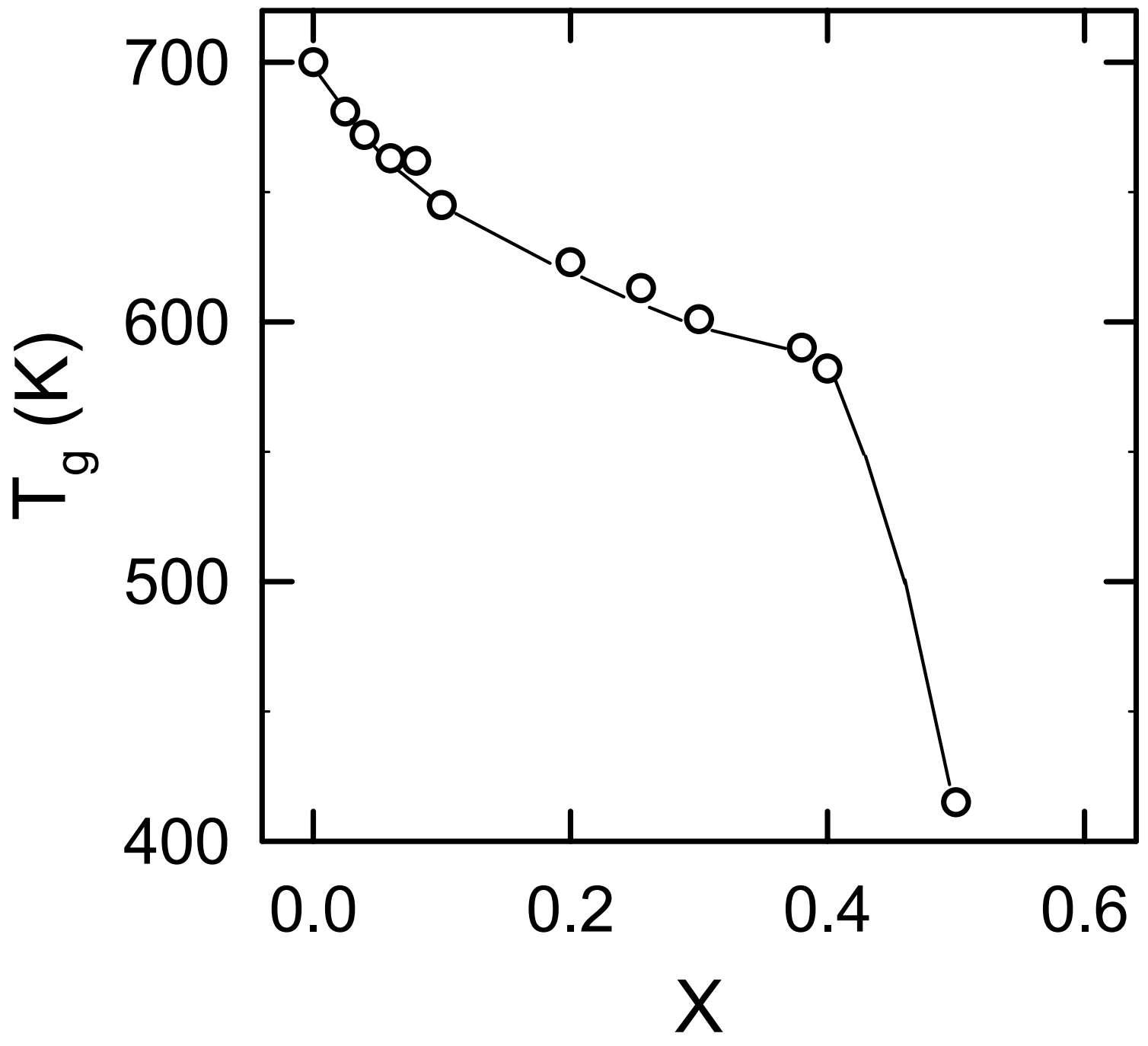

Fig. 3. The glass transition temperature $T_g$ in $Sn_xGe_{1-x}Se_2$ glass alloys. At the predicted critical composition x = 0.40 the network softens dramatically. Angular constraints are weak and so ignored at the Sn ions but are strong and intact for Ge and Se ions.

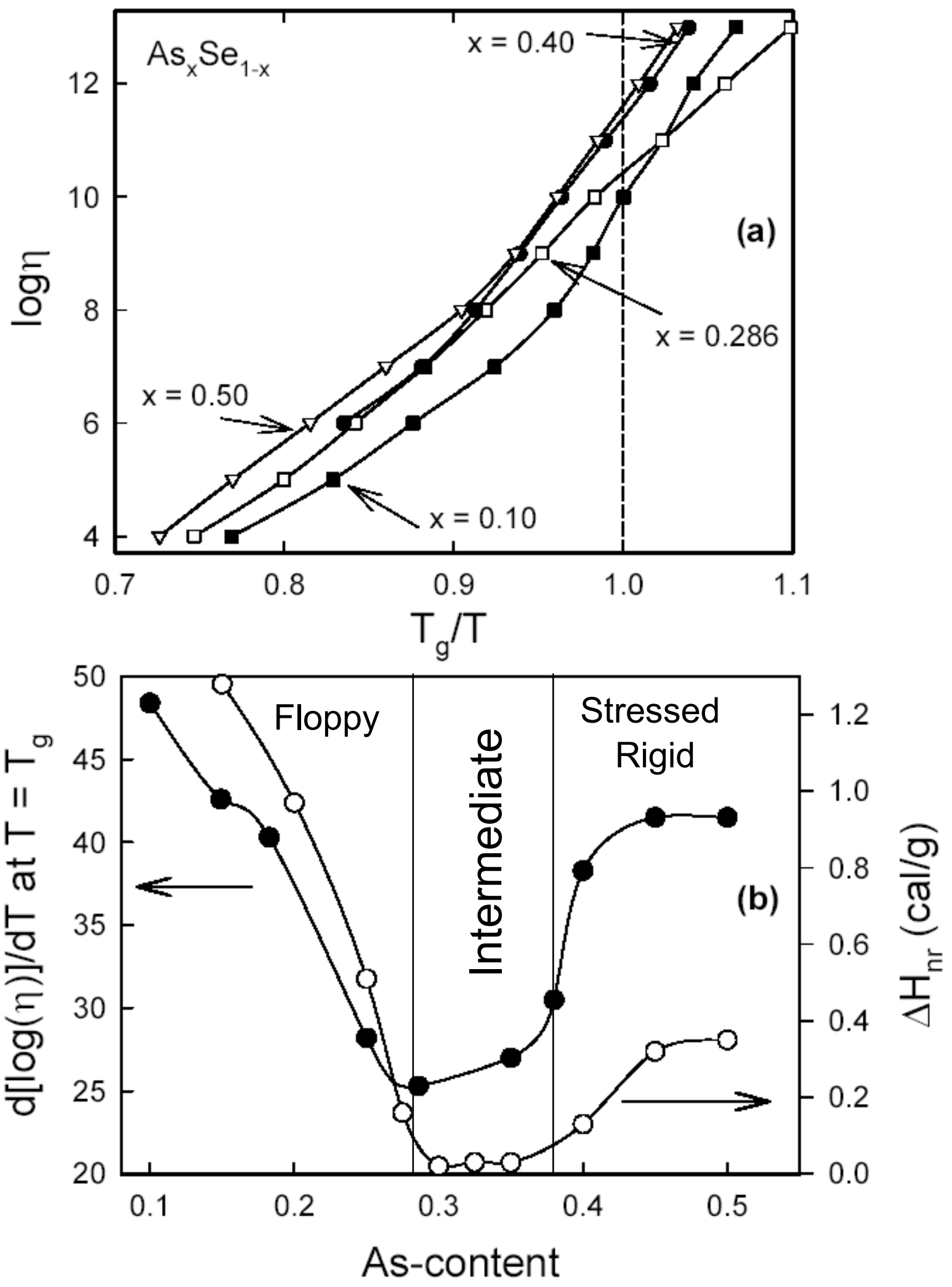

Fig.4. (a) Viscosity η(T) of four $As_xSe_{1-x}$ glass melts as a function of temperature; a rare example of a glass alloy system for which such complete viscosity data are available. (b) Comparison of dlogη/dT at $T_g$ with the nonreversible enthalpy $\Delta H_{nr}$ of the glass transition as functions of x, as measured by Modulated Differential Scanning Calorimetry. The latter are available for many alloys. Note the different base lines. The former drops by a factor of two in the window, whereas the latter drops by more than a factor of 10 (nearly to zero). Thus the reversibility window is resolved much more accurately by the modern method (MDSC), enabling its identification, and giving accurate estimates of its compositional boundaries.

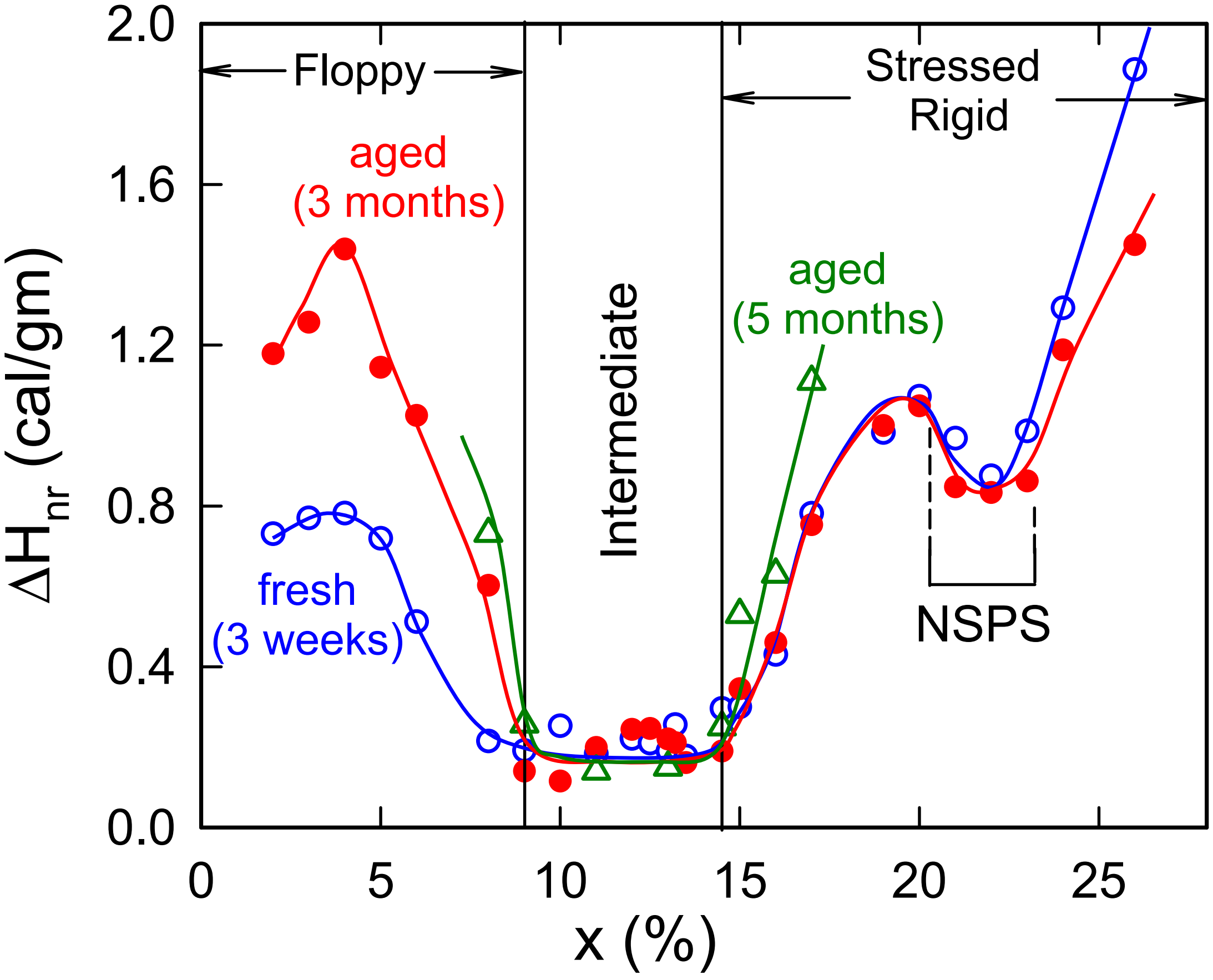

Fig. 5. The special nature of glassy networks in the non-reversibility window is brought out in spectacular fashion by studying the aging of $\Delta H_{nr}$ at temperatures well below $T_g$ for periods of order months. Within the window there is very little change, but outside the window aging causes large increases in $\Delta H_{nr}$. Nanoscale Phase Separation of $P_4Se_4$ and $P_4Se_3$ molecules from the backbone results in the feature labeled NSPS. The alloy studied here is $Ge_xP_x\ Se_{1-2x}$.

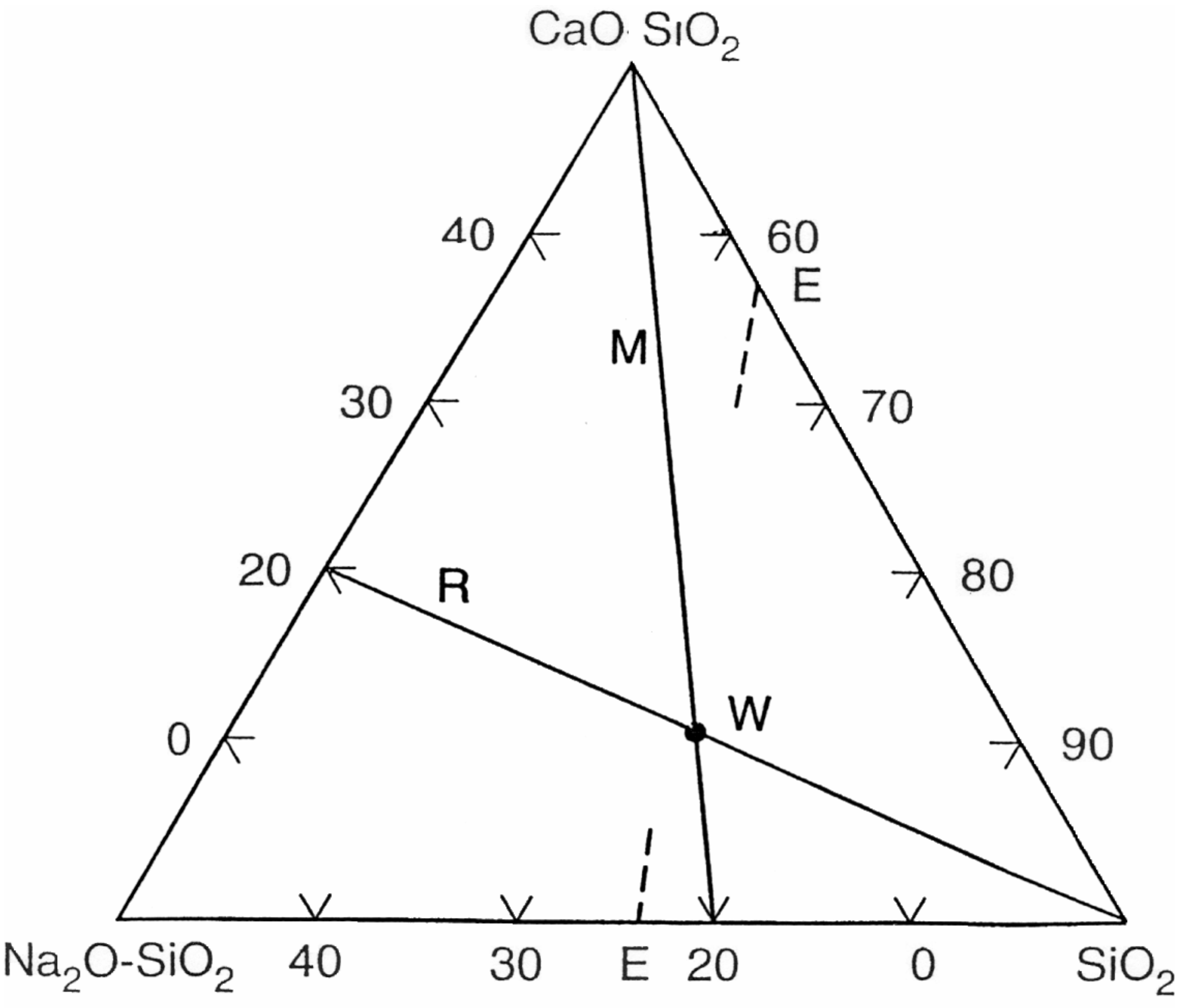

Fig. 6. A ternary phase diagram for silica-soda-lime glass alloys. The line labeled M is the Mean-field line for a global stress-free network, while the line labeled R represents the local topological condition of constant Ring size. The dotted line segments labeled E are eutectics. The lines M and R intersect at W, which is the composition of Window glass.

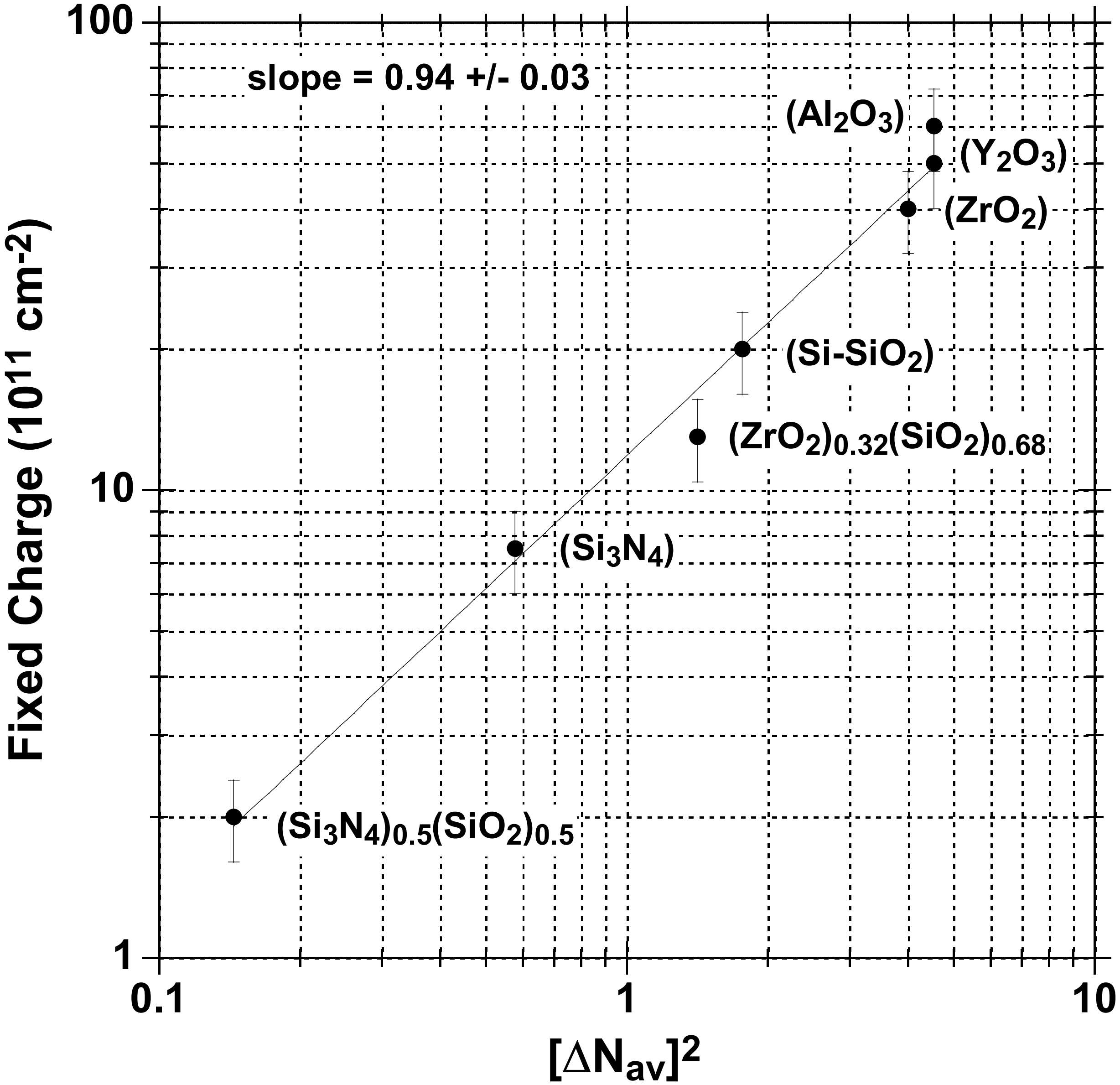

Fig. 7. Fixed charge (trap density) at Si-dielectric interfaces for dielectrics of technological interest. Here $N_{av}$ is the average valence number of the dielectric and $\Delta N_{av} = N_{av} - 2.67$

.

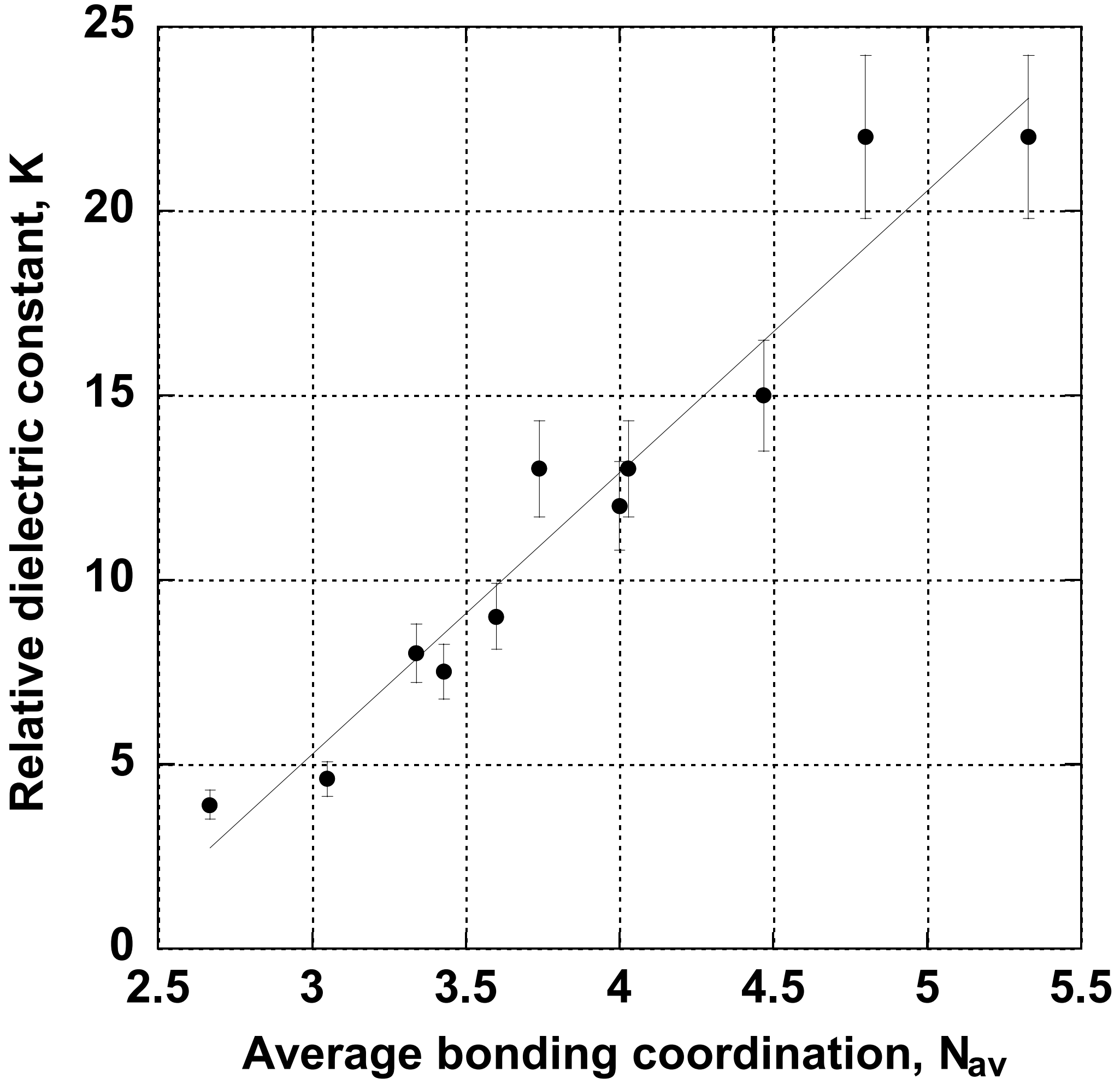

Fig. 8. Dielectric constant k for many Zr- and Hf-doped dielectrics relative to silica. Here $N_{av}$ is the average valence number of the dielectric.

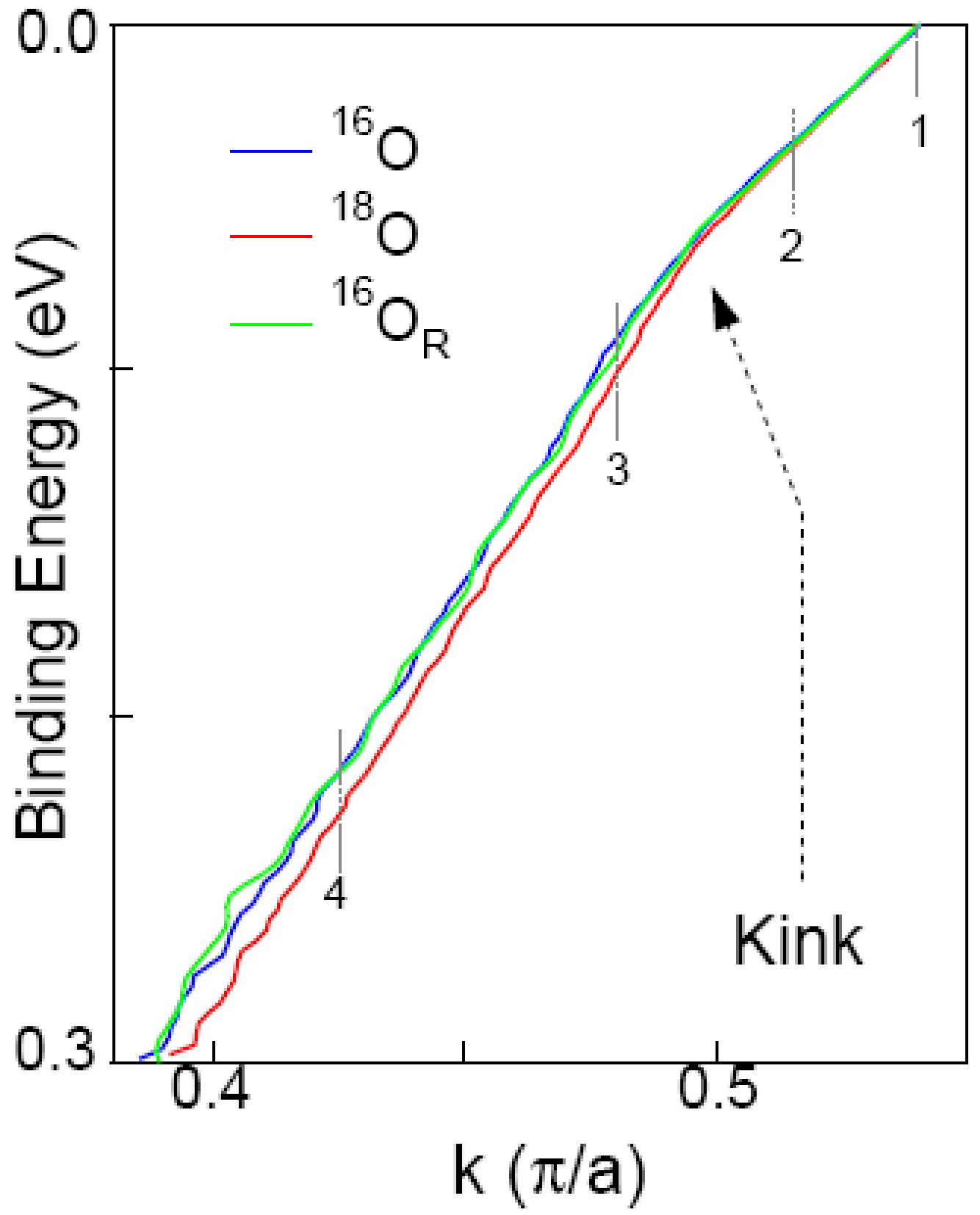

Fig. 9. Isotope shifts measured by angle-resolved photoemission. There is a kink in the dispersion curve near 70 meV; above this kink (points 1 and 2, closer to the Fermi energy, set equal to 0) the isotope shift is small, below it (points 3 and 4) the shift grows linearly. These shifts are inexplicable using polynomial methods (for example, conventional polaron theory).

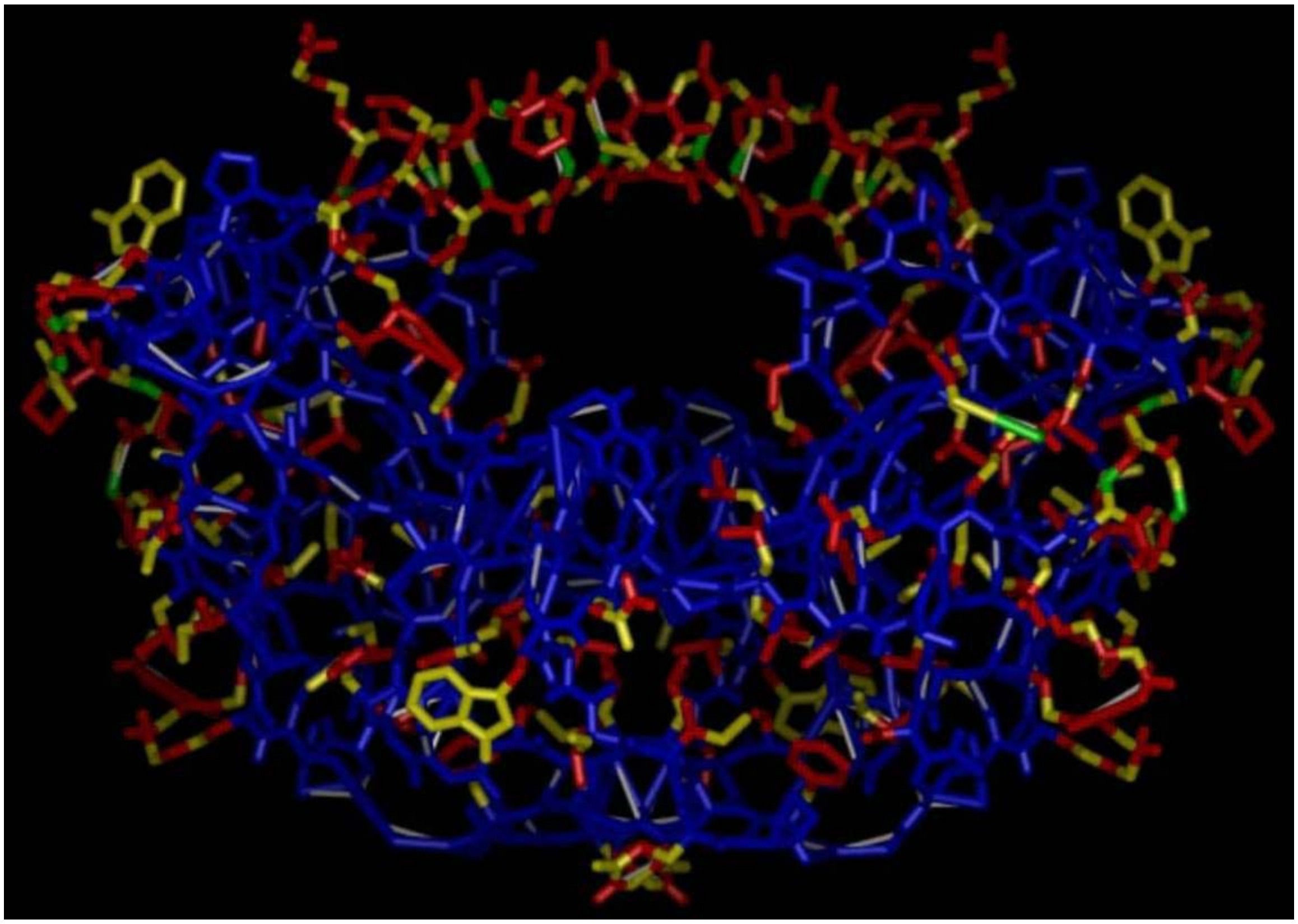

Fig. 10. The protein HIV protease is a homodimer containing ~ 200 amino acid residues. It can be separated into flexible parts (shown in reds, yellows and greens) and a rigid core (shown in blue) using algorithms based on the pebble game[12]. The function of HIV protease can be inhibited by pinning the flaps at the top by the binding of an appropriate ligand.